\documentclass[preprint,authoryear,12pt]{elsarticle}

\usepackage{graphicx}
\usepackage{amssymb}

\usepackage{amsmath} 
\usepackage{hyperref}

\journal{arXiv}

\begin{document}

\begin{frontmatter}



\title{Interictal intracranial EEG for predicting surgical success: the importance of space and time}


\author[label1,label2,label3,label4]{Yujiang Wang}
\author[label1,label2]{Nishant Sinha}
\author[label1]{Gabrielle M Schroeder}
\author[label1]{Sriharsha Ramaraju}
\author[label3]{Andrew W McEvoy}
\author[label3]{Anna Miserocchi}
\author[label3]{Jane de Tisi}
\author[label3]{Fahmida A Chowdhury}
\author[label3]{Beate Diehl}
\author[label3]{John S Duncan}
\author[label1,label2,label3,label5]{Peter N Taylor}

\address[label1]{School of Computing Science, Newcastle University, UK}
\address[label2]{Institute of Neuroscience, Newcastle University, UK}
\address[label3]{Institute of Neurology, Queen Square, London WC1N 3BG, UK}
\address[label4]{yujiang.wang@newcastle.ac.uk}
\address[label5]{peter.taylor@newcastle.ac.uk}

\begin{abstract}
Predicting post-operative seizure freedom using functional correlation networks derived from interictal intracranial EEG has shown some success. However, there are important challenges to consider. 1: electrodes physically closer to each other naturally tend to be more correlated causing a spatial bias. 2: implantation location and number of electrodes differ between patients, making cross-subject comparisons difficult. 3: functional correlation networks can vary over time but are currently assumed as static.
 
In this study we address these three substantial challenges using intracranial EEG data from 55 patients with intractable focal epilepsy.  Patients additionally underwent preoperative MR imaging, intra-operative CT, and post-operative MRI allowing accurate localisation of electrodes and delineation of removed tissue.
 
We show that normalising for spatial proximity between nearby electrodes improves prediction of post-surgery seizure outcomes.  Moreover, patients with more extensive electrode coverage were more likely to have their outcome predicted correctly (ROC-AUC $>$0.9, p$<<$0.05), but not necessarily more likely to have a better outcome. Finally, our predictions are robust regardless of the time segment.
 
Future studies should account for the spatial proximity of electrodes in functional network construction to improve prediction of post-surgical seizure outcomes. Greater coverage of both removed and spared tissue allows for predictions with higher accuracy.
\end{abstract}


\end{frontmatter}



\section{Introduction}
\label{intro}

Surgery is an effective treatment for epilepsy, with over half of patients achieving outcomes of postoperative seizure freedom \citep{de_tisi_long-term_2011}.  For patients not seizure free after surgery, a possible explanation is the incomplete removal of the epileptogenic zone, defined as the area of cortex that is indispensable for seizure generation \citep{rosenow_presurgical_2001}. More recently, the concept of the epileptogenic ‘network’ has emerged, recognising multiple brain regions and connections between them to be responsible for generating seizures \citep{bartolomei_defining_2017,spencer_neural_2002}. Identification of the epileptogenic network in each patient is extremely challenging, as removal of multiple brain regions and connections between them may lead to seizure freedom. Several recent studies have highlighted the potential network properties that are specific to the epileptogenic network and used various network properties to predict post-operative outcome \citep{richardson_large_2012,englot_global_2015,munsell_evaluation_2015,morgan_magnetic_2017,taylor_impact_2018}.

Functional networks inferred using intracranial EEG (iEEG) have received considerable attention in this context. These functional networks measure iEEG signal similarity as connection strength between iEEG channels. Indeed, studies using iEEG-derived networks have demonstrated their value for predicting patient outcomes when using ictal \citep{wilke_graph_2011,burns_network_2014,goodfellow_estimation_2016,park_granger_2018,yang_localization_2018} and inter-ictal data \citep{palmigiano_stability_2012,sinha_silico_2014,zweiphenning_high_2016,sinha_predicting_2017,tomlinson_interictal_2017,shah_high_2019}.

The potential of using (only) interictal data is particularly attractive in a clinical setting \citep{korzeniewska_ictal_2014,shah_high_2019}, but despite promising results from previous studies, several challenges remain.  First, electrodes that are physically closer are also more likely to be highly correlated \citep{lagarde_interictal_2018,betzel_structural_2019}. This means that functional networks derived from iEEG are always dependent on the spatial layout of electrodes, which differs from patient to patient. Second, the individualised configuration of intracranial electrodes also means that different number of electrodes are sampling tissue that is ultimately removed or spared by surgery. This makes comparison of network properties across patients difficult. Third, although iEEG functional networks fluctuate over time \citep{kuhnert_long-term_2010,geier_time-dependent_2015}, it is not currently known if these fluctuations affect their use to predict surgical outcome. To enable the translational use of interictal functional iEEG networks, it is crucial to answer these questions.

\section{Materials \& Methods}
\subsection{Patients and interictal EEG and MRI preprocessing}
This retrospective study analysed data from 55 patients with refractory focal epilepsy from the National Hospital for Neurology and Neurosurgery (NHNN) who had intracranial EEG (iEEG) followed by resection and clinical follow-up of at least 12 months. iEEG data was acquired using a mixture of grid, strip, and SEEG recordings. The iEEG data was anonymised and exported, then analysed under the approval of the Newcastle University Ethics Committee (2225/2017). Patient metadata is shown in supplementary material S1

The iEEG data analysed consisted of continuous one-hour segments of interictal EEG that were at least two hours away from seizures. Artefactual channels were removed by visual inspection, and all remaining channels were subsequently re-referenced to common average. Each channel was then notch filtered at 50 Hz (iir filter with Q factor 50, 4th order zero phase lag) and bandpass filtered (Butterworth 4th order zero phase lag) between 1-70 Hz.

To delineate the iEEG electrodes that overlapped with the subsequently surgically removed tissue, we first mapped the spatial position of the iEEG electrodes to the space of the pre-operative T1w MRI using intraoperative MRI and CT in a semi-automated fashion \citep{hamilton_semi-automated_2017}. We also manually delineated the surgically removed tissue in space of the pre-operative MRI using rigid body registration of the post-operative T1w MRI to the pre-operative T1w MRI \citep{taylor_impact_2018}. Any iEEG electrode contact that was within the 5mm distance of surgically removed tissue was deemed as a “removed electrode”. All others were marked as “spared electrodes”. For one patient (IDP 851), a post-operative MRI was unavailable and the surgery report from the clinical team was therefore used to identify the resected electrodes. The procedure is summarised schematically in Fig 1A.
\subsection{Functional networks derivation and network quantification}
Functional brain networks were derived from the one-hour iEEG segments. We applied Pearson correlation to 2 second sliding windows (without overlap) of the broad band (1-70 Hz) iEEG data and subsequently averaged the correlation matrices over all windows to obtain one functional network (average correlation) matrix per subject. 

To quantify the network properties of each node we use the node strength of the network, which is a measure of the total level of correlation of a node. It has been suggested that this quantity is indicative of epileptogenic tissue, and we show in supplementary material S3 that it recapitulates a quantity derived from a dynamical model of epileptogenic tissue we previously suggested \citep{sinha_predicting_2017}. 
To quantify if the node strength of removed electrodes differed from the node strength of the spared electrodes, we used the area under the receiver-operator curve (AUC), which is equivalent to the normalised non-parametric Mann-Whitney U statistic. We chose this measure because it is based only on the rank order of the node strength, and thus robust to outliers and non-normal distributions in node strength. In the following we will term this measure $D_{RS}$ which stands for the Distinguishability of the Removed node strength vs. the Spared node strength and is a single value per patient.
This procedure is summarised schematically in Fig 1B.

\begin{figure}
\begin{centering}
\includegraphics[width=14cm]{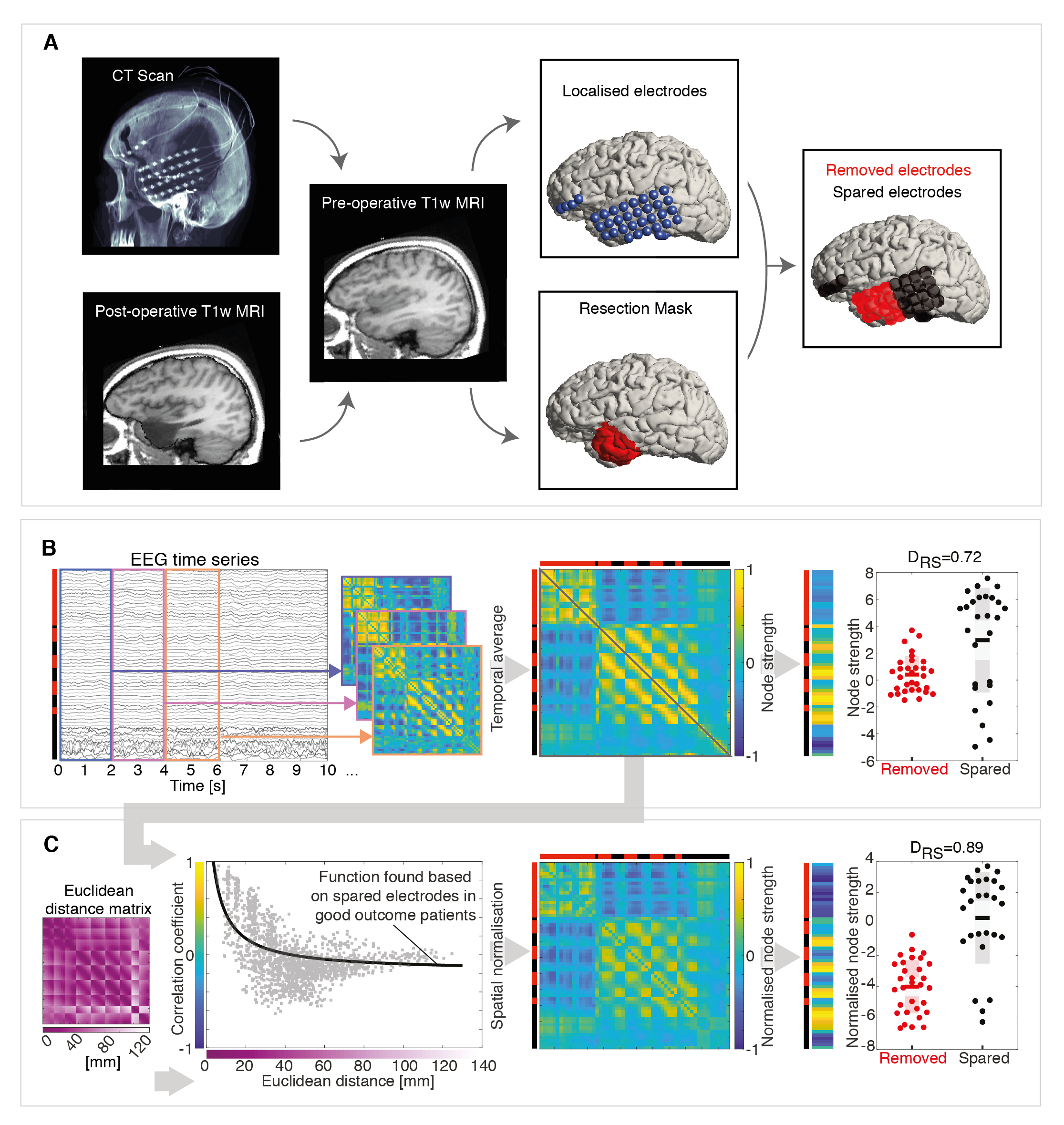}
\par\begin{tiny}
\end{tiny}\end{centering}
\caption{\textbf{Summary of the processing pipeline. }\textbf{A} CT and post-operative T1 weighted MRI scans were linearly registered to the pre-operative T1 weighted MRI scan.  Electrodes, shown in blue, were localised from the registered CT and marked with respect to the cortical surface.  A mask was additionally manually drawn (shown in red and projected to the cortical surface for visualisation) to cover those areas removed by surgery. Electrodes located within 5 mm of the volumetric surgery mask were then identified as ‘removed’ (shown in red) and all others as identified as ‘spared’ (shown in black). The Euclidean distance between each electrode is also calculated in millimetres. \textbf{B} Two second non-overlapping correlation matrices are computed from the EEG time series, and their mean (temporal average) matrix calculated.  By summing the rows of the temporal average matrix, the node strength was calculated.  The difference in node strengths for removed and spared electrodes was then computed as the $D_{RS}$ measure, with 1 indicating perfect separation of the removed vs. spared tissue and 0.5 indicating no separation. \textbf{C} To derive a spatially normalised temporal average matrix we applied a spatial regression which was pre-calculated based on spared electrodes from good outcome patients. After applying the regression, we retained the residuals as the spatially normalised temporal average matrix, which allows to calculate a normalised node strength and $D_{RS}$ value.  The pipeline was applied to each patient.  } \label{fig:fig1}
\end{figure}

\subsection{Spatial normalisation}
The fact that electrodes that are closer together in space are more likely to be correlated introduces a bias in the functional networks that depends on each subject’s implantation. We therefore applied a spatial normalisation to reduce this bias. We used spared electrodes from good outcome (ILAE class 1) patients only - which represents signals from non-epileptogenic tissue - as a baseline to establish how correlation coefficients change as a function of spatial (Euclidean) distance. Once this baseline function was found, all correlation coefficients in the functional network matrices were normalised by computing the residuals to this baseline function. The spatially-normalised correlation coefficients now indicate the extent of correlation between two electrodes that is beyond what is expected due to their spatial proximity to each other. 
This procedure is summarised schematically in Fig 1C.

\subsection{Spatial coverage}
Due to clinical need, the spatial coverage of the implanted electrodes differs from patient to patient. Thus, the spatial sampling of the surgically removed/spared tissue differs in terms of total number of electrodes and amount of tissue removed. To account for these differences, we obtained the number of removed electrode contacts and the number of spared electrode contacts in each patient. We then successively excluded subjects from our analysis based on the minimum number of removed and spared electrode contacts ($n_x$) to observe the effect of spatial coverage on our results. For example, at $n_x=1$, all 55 subjects are included. At $n_x = 20$, only subjects who have at least 20 removed electrode contacts and at least 20 spared electrode contacts are retained for analysis, which in our case consists of 27 subjects. In our analysis we scan the $n_x$ parameter from 1 to 40 and report the results for each value.  The rationale behind this measure is that larger $n_x$ values can be interpreted as providing a better network representation and sampling of both removed and spared networks. It then follows that if we have captured the network better, our discrimination between outcome groups should improve.

\subsection{Temporal variability}
A longstanding open question in the field is if the temporal variability of the interictal functional networks \citep{kuhnert_long-term_2010,geier_time-dependent_2015} affect its ability to delineate epileptogenic tissue.

We first addressed the question of time scale by dividing the one-hour iEEG segment into several smaller non-overlapping segments (segments of length 4 seconds, 10 seconds, 20 seconds, 40 seconds, 1 minute, 3 minutes, 6 minutes, and 10 minutes) and measuring their performance when repeating the same analysis. Specifically, for each segment, we apply a 2s non-overlapping windows to create an average functional network matrix for this segment. For example, the 40 second segment functional network matrix is created from 20 windows, whereas the 10 second segment is generated from only 5 windows.

We next investigated the performance of two other separate one-hour segments from the same subject (at least 2 hours away from seizures, and at least 4 hours away from the other iEEG segments). In some patients, it was not possible to find such a second or third one-hour segment, and thus the number of patients decreases slightly (53 subjects had a second segment, and 51 subjects had a third segment).

All analyses were performed independently on each segment; all the steps including spatial regression onwards were performed for each segment without knowledge of the other segments.
\subsection{Statistical analysis of relationship to surgical outcome}
To investigate if $D_{RS}$ contains useful information to explain post-surgical outcomes, we tested if $D_{RS}$ is different between good (ILAE class 1) and poor outcome (ILAE class 2 and above) patients. We measured the area under the receiver-operator curve (AUC) as a main indicator, where AUC equal to 1 shows that $D_{RS}$ can fully distinguish good and poor outcome patients. Conversely, an AUC of 0.5 indicates that $D_{RS}$ cannot distinguish between outcome groups. We also tested for the statistical significance of the AUC by performing the ranksum test between the good and poor outcome patients for the $D_{RS}$ measure. We obtained 95\% confidence intervals of the AUC based on a logit transformation \citep{qin_comparison_2008}. Supplementary material S2 shows equivalent results using a cross validated AUC.

\subsection{Data availability}
We will make all functional network matrices and analysis code underlying all result figures available upon acceptance of the manuscript.

\section{Results}
In this study, we compare network properties (in particular node strength) of interictal iEEG functional networks between the surgically removed and spared tissue to predict surgical outcome (seizure freedom) in individual patients.  We will specifically address the following three questions: (1) Does spatial normalisation of functional networks increase the ability to distinguish between outcome groups? (2) Does increased coverage of removed and spared tissue lead to increased distinction between outcome groups? (3) Does the choice of timescale or time point affect the ability to distinguish between outcome groups? 
\subsection{Spatial normalisation of interictal functional networks improves distinction between outcome groups}
We first investigate if node strength computed from raw, spatially un-normalised networks discriminates between outcome groups. Figure 2A,B (upper panels) shows the node strength computed for two example patients which are then used to calculate the $D_{RS}$ value.  This single $D_{RS}$ value measures the difference in node strength between removed and spared electrodes in an individual. Figure 2C shows the $D_{RS}$ value for all 55 patients in our study and we find no substantial difference in $D_{RS}$ value between outcome groups (Fig. 2D).
\begin{figure}
\begin{centering}
\includegraphics[width=14cm]{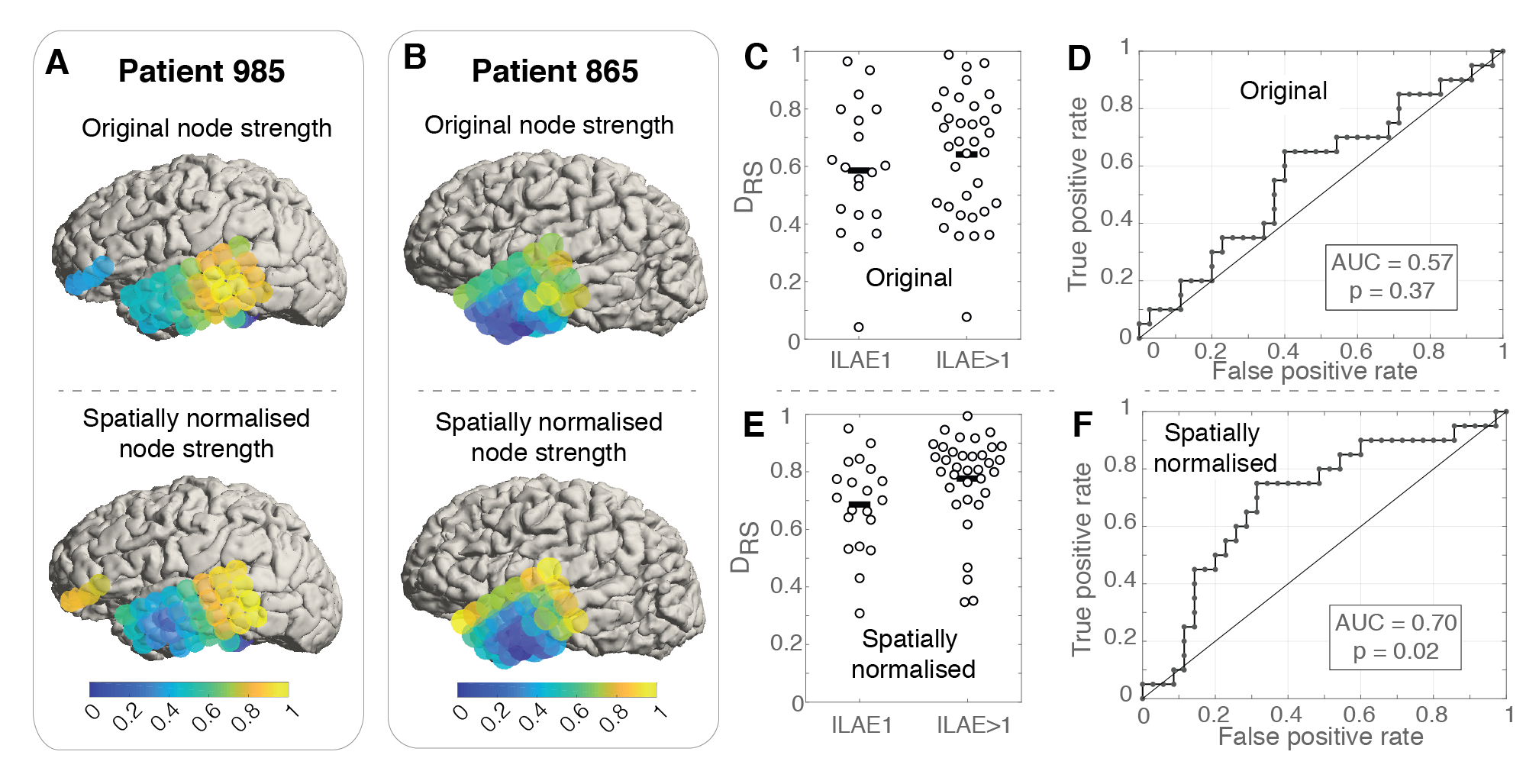}
\par\begin{tiny}
\end{tiny}\end{centering}
\caption{\textbf{Spatial normalisation improves the discrimination between outcome groups. }\textbf{A,B} Node strength before and after spatial normalisation for two example patients. \textbf{C} ILAE outcome groups show similar $D_{RS}$ values, meaning that the difference in node strength between removed and spared nodes does not explain outcome. Each dot is an individual patient. \textbf{D} Receiver operating curve for the data presented in panel C shows poor discrimination between outcome groups. \textbf{E} After spatial normalisation patient groups show significant differences in their $D_{RS}$ values, which can discriminate outcome groups with a ROC AUC of 0.7 (F).  } \label{fig:fig2}
\end{figure}
Given that electrodes which are more spatially proximal are more likely to have higher functional connectivity, we next sought to determine if normalising for this effect increases discrimination between groups using the $D_{RS}$ measure. To this end we used a null model for spatial normalisation which accounts for the spatial proximity of electrodes (see Methods). The normalisation subsequently impacts the node strength (c.f. Fig. 2A,B upper and lower panels), and in our data it improves the distinction between outcome groups (compare Fig 2C,D with E,F). For our remaining analysis we will therefore use spatially normalised functional networks.

\subsection{Increased coverage of removed and spared tissue improves distinction between outcome groups }
Spatially undersampling networks can directly lead to changes in the estimated network properties \citep{conrad_how_2019}, and thus we investigated the impact of spatial sampling on our ability to distinguish outcome groups. Fig. 3A shows two example patients: one has a large number of electrode contacts located in both removed and spared tissue, but the other one only has eight electrode contacts located in the spared tissue. Therefore, these two patients are not directly comparable in terms of the network properties of their spared tissue. To account for this issue, we successively excluded patients with a low $n_x$, which is their minimum number of electrodes in both spared and removed tissue. As we increase coverage of removed and spared tissue ($n_x$) the distinction between outcome groups in terms of AUC values becomes clearer (Fig. 3B). For example, at $n_x=20$ (we only use the subset of patients that have at least 20 electrodes spared and removed tissue), 27 patients remain for the analysis, and we find outcome class is distinguishable with an AUC=0.91 (Fig. 3C). Note also that the proportion of good vs. poor outcome patients does not change substantially over $n_x$ (Fig. 3B bottom panel orange line). Cross-validated AUCs follow a similar trend of increasing AUC for greater coverage (supplementary information S2).

\begin{figure}
\begin{centering}
\includegraphics[width=10cm]{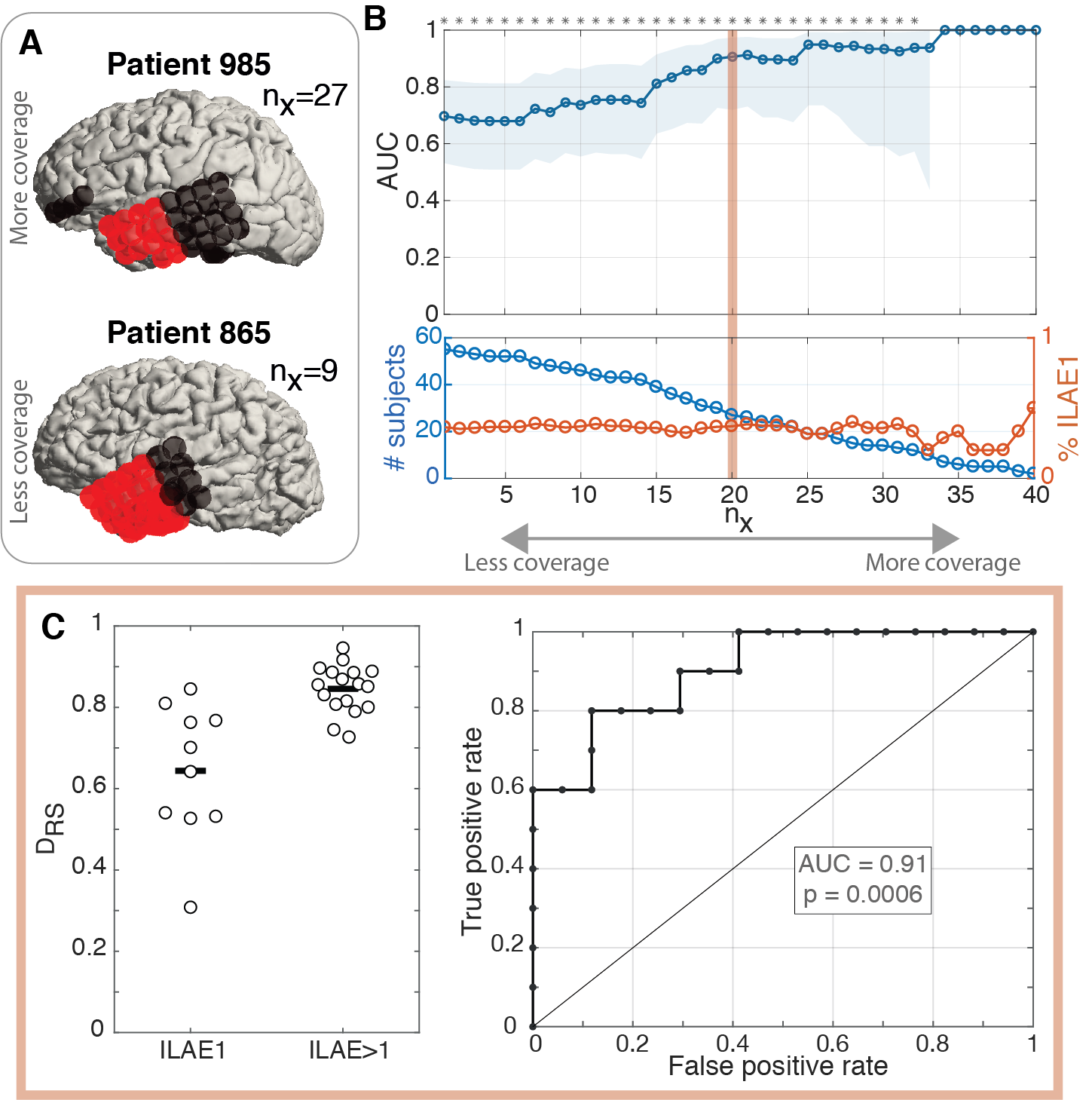}
\par\begin{tiny}
\end{tiny}\end{centering}
\caption{\textbf{Greater coverage of spared and removed networks leads to greater discrimination between outcome groups. }\textbf{A} Removed (red) and spared (black) electrodes are shown for two example patients. Patient 985 has greater sampling of spared tissue (27 electrodes) than patient 865 (9 electrodes). \textbf{B} Top: Scanning AUC (blue line) over $n_x$, where at each $n_x$ value, only patients with at least $n_x$ electrodes in removed and spared tissue are included in the analysis. Shaded blue area indicates the 95\% confidence interval for the AUC. For high $n_x$ too few subjects remained for analysis to obtain a confidence interval. Bottom: Showing the number of patients included (blue line) and the percentage of good outcome (orange line) patients for each $n_x$ value.\textbf{ C} At a value of $n_x=20$, $D_{RS}$ values between outcome groups, and ROC curve are shown.} \label{fig:fig3}
\end{figure}
\subsection{Interictal functional networks fluctuate over time, but these fluctuations do not affect the distinction between outcome groups}
For practical applications using interictal iEEG functional networks to delineate epileptogenic tissue, it is important to understand if and how our results change if different underlying data are used.  Specifically, we investigate if networks generated from segments of different duration and from different time points affect our main findings. We systematically scanned segments of shorter durations and measured their ability to distinguish outcome groups in our cohort (Fig. 4A,B). Typically, a 10 second segment does not perform significantly worse than a 1 hour segment over all $n_x$. For all segments of all lengths, their AUCs lie largely within the 95\% confidence interval of the AUC for the one-hour segment (Fig. 4A,B). We do however note that the AUC varies more from segment to segment for shorter segments ($<=$10s), indicating that consistency of results may drop for short segments.

\begin{figure}
\begin{centering}
\includegraphics[width=14cm]{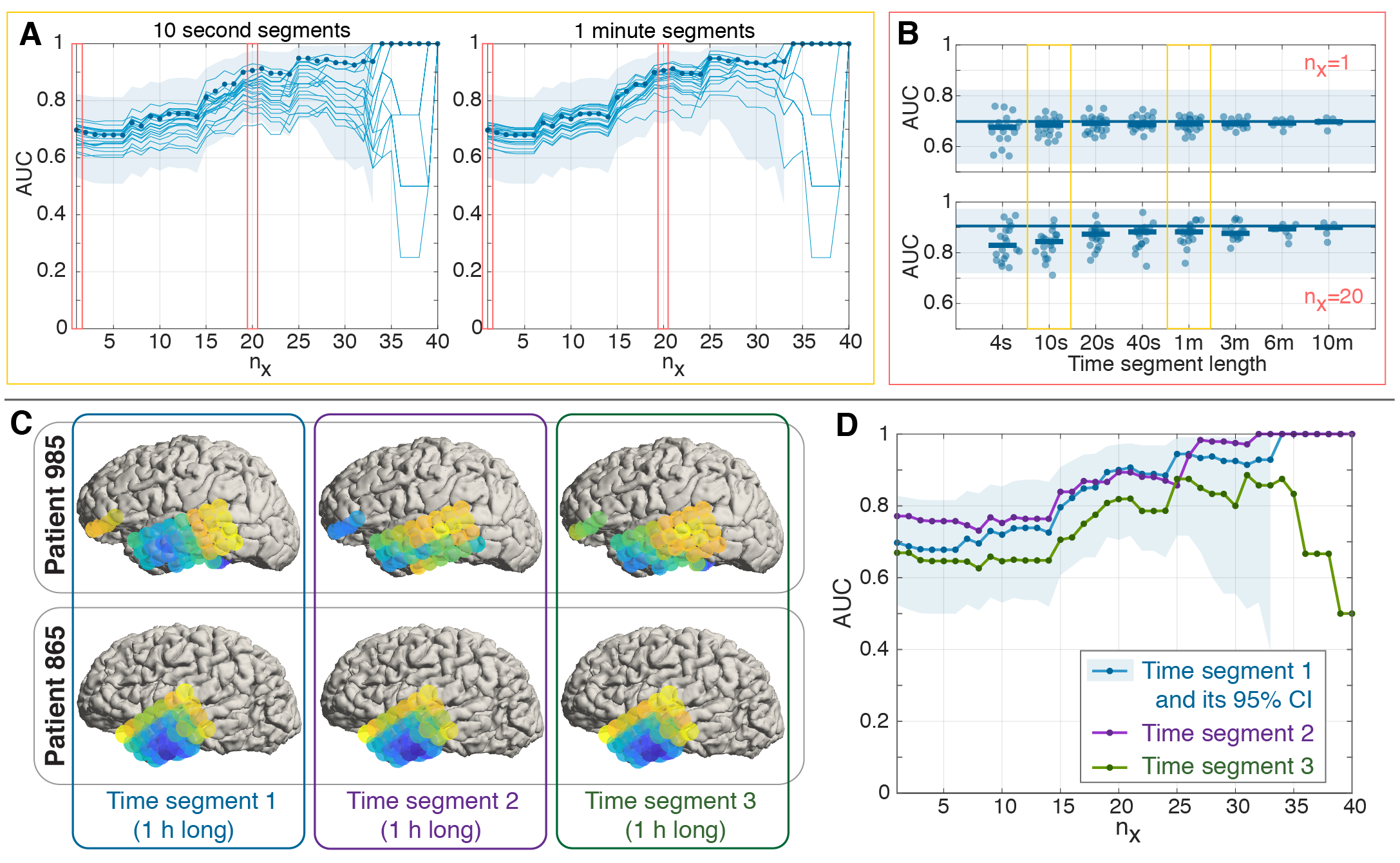}
\par\begin{tiny}
\end{tiny}\end{centering}
\caption{\textbf{Consistent results for networks sampled at different times.} \textbf{A} Left: AUCs for 20 separate ten second segments are shown as individual thin blue lines. AUC of the 1h segment (thick blue line) and 95\% confidence intervals (shaded blue area) are reproduced from figure 3B for reference. Right: same for 20 separate one minute segments. \textbf{B} AUC for segments of different length shown for $n_x=1$ and $n_x=20$.  The horizontal blue line (shaded area) indicates the AUC (confidence interval) for the one-hour segment for reference. Thick horizontal blue bars indicate the mean AUC of all segments of a particular length. \textbf{C} Node strength for two example patients for three separate one-hour time segments. Time segment 1 panels are identical to those in the lower panels of figure 2A,B. \textbf{D} Thick blue line and shaded area are from the first one-hour segment and reproduced from figure 3B for reference. Purple and green lines indicate the AUC for the second and third one-hour segments, respectively. Note that all functional networks of all segments were derived by averaging over correlation matrices from 2s non-overlapping windows in this figure. I.e. the window size to obtain correlations of the time series stayed constant in all segments.} \label{fig:fig4}
\end{figure}
To test if a different one-hour time segment would change our AUCs substantially, we repeated the same analysis for two different one-hour time segments that are at least four hours away from any other segment in each patient. Fig. 4C shows the node strength of two example subject over all three one-hour segments, and although there are some variations, the gross spatial pattern remains stable. This consistency is also reflected in the AUCs distinguishing outcome groups (Fig. 4D), where the two new segments lie within the confidence interval of the original segment.

\section{Discussion}

Interictal iEEG network-based approaches to predict seizure freedom after surgery and to identify epileptogenic tissue have attracted interest in recent years.  However, major issues regarding spatial bias, incomplete coverage, and temporal stability have remained relatively unexplored. Our study makes three key contributions in this regard.  First, we found that spatial normalisation substantially increases the discrimination between outcome groups.  Second, we found that increased coverage of removed and spared networks leads to greater distinction between outcome groups, but not necessarily better outcomes.  Third, we show that our results are consistent for a wide range of time scales from minutes up to hours. Our work confirms that interictal iEEG network analysis holds value for predicting seizure freedom after surgery.

Spatial normalisation of functional networks is a natural way to measure signal similarity relative to a baseline. The need to establish baselines derived from “healthy” tissue for iEEG has also been recognised by others \citep{frauscher_atlas_2018,betzel_structural_2019}. Besides spatial information, white matter pathways and shared gene expression between regions also explain functional relationships \citep{betzel_structural_2019}. In future, these and other variables could also be included to normalise functional networks in iEEG to enhance the detection of the pathological aspects.  

Intracranial EEG only samples specific subnetworks in the brain, which can vary widely between patients. It is clear that such subnetworks do not necessarily have the same properties as the whole-brain network. Recent analyses demonstrated that even leaving out one node from iEEG functional networks can dramatically change their network properties \citep{conrad_how_2019}. The implication is that the property of the epileptogenic tissue/network may change depending on the subnetwork sampled, which may explain some conflicting results in the literature \citep{zijlmans_changing_2019}. Thus, the restricted spatial sampling inherent in iEEG is a natural limitation in the context of functional networks, and we showed that it directly impacts upon how informative the functional networks are for distinguishing outcome groups. Other studies using different patient cohorts (with different implantation strategies) may therefore achieve better or worse AUCs as a direct consequence of the coverage of the patients in the study. Finally, it is important to note that in our study, increased coverage (higher $n_x$) leads to increased discrimination between outcome groups, but does not necessarily lead to better outcomes. Future studies combining scalp EEG or other recording modalities with iEEG may reveal how the iEEG subnetwork can be related to whole-brain networks to improve localisation.

Temporal fluctuations of iEEG functional networks are well-studied during epileptic seizures and the pre-ictal periods \citep{burns_network_2014,campo_degenerate_2018,schroeder_slow_2019}; however, the interictal iEEG functional networks are often treated as static \citep{palmigiano_stability_2012,sinha_predicting_2017,shah_high_2019} and in contrast to the ictal networks, it is suggested that interictal networks are stable over time \citep{kramer_emergence_2011,chapeton_stable_2017}. Rather than determining stability or fluctuations of interictal functional networks as such, we asked the simpler question of whether the time scale or time point matters for discriminating between outcome groups. In our cohort, the time scale and time point did not dramatically impact our results. However, this result should not be directly interpreted as evidence for stability of the interictal functional networks. Future work should investigate what aspects of interictal iEEG are variable/static, which may also highlight the causal link between interictal EEG and epileptogenic tissue.

Taken together, our results support the use of interictal intracranial EEG networks for predicting surgical outcome and provide considerations and practical solutions for its clinical use. Future studies should investigate the generalisability of the approach across multiple clinical sites, and assess the combined use with noninvasive whole-brain approaches, such as high density scalp EEG.

\section{Acknowledgements}
We thank Gerold Baier, Louis Lemieux, Richard Rosch, and members of the CNNP lab (\url{www.cnnp-lab.com}) for discussions on the analysis and manuscript. We thank Catherine Scott, Roman Rodionov, and Sjoerd Vos for helping with data organisation. B.D. receives support from the NIH—National Institute of Neurological Disorders and Stroke U01-NS090407 (The Center for SUDEP Research) and Epilepsy Research UK. P.N.T. and Y.W. gratefully acknowledge funding from Wellcome Trust (208940/Z/17/Z and 210109/Z/18/Z). The authors declare no conflict of interest. The funders played no role in the design of the study.
\section{Author contributions}
Y.W. and P.N.T conceived the idea, developed the methods, and wrote the code. B.D. oversaw clinical acquisition and annotation of the NHNN patient EEG data.  J.D. oversaw clinical acquisition and annotation of the UCLH patient MRI data.  JdT. oversaw clinical annotation of the NHNN patient metadata. F.A.C. contributed to clinical acquisition of UCLH patient EEG data. Y.W. and P.N.T organised the data and performed the analysis. Y.W. and P.N.T. drafted the manuscript and produced all figures. All authors participated in critically reviewing and revising the manuscript.


\bibliographystyle{elsarticle-harv}
\bibliography{references}


\end{document}